\documentclass[12pt]{article}

\usepackage{amsmath,amssymb,cite}
\usepackage{graphicx}

\makeatletter
 
  \@addtoreset{equation}{section}
 \makeatother

\newcommand{\tr}{\mathrm{tr}}
\newcommand{\Tr}{\mathrm{Tr}}

\newcommand{\del}{\partial}

\newcommand{\n}{\nonumber \\}

\newcommand{\cL}{\mathcal{L}}

\def\XXint#1#2#3{{\setbox0=\hbox{$#1{#2#3}{\int}$} 
\vcenter{\hbox{$#2#3$}}\kern-.5\wd0}}

\allowdisplaybreaks

\begin{document}

\setlength{\oddsidemargin}{0cm}
\setlength{\baselineskip}{7mm}

\begin{titlepage}
\renewcommand{\thefootnote}{\fnsymbol{footnote}}
\begin{normalsize}
\begin{flushright}
\begin{tabular}{l}
KUNS-2245\\
RIKEN-TH-179\\
December 2009
\end{tabular}
\end{flushright}
  \end{normalsize}

~~\\

\vspace*{0cm}
    \begin{Large}
    \begin{bf}
       \begin{center}
         {Large $N$ reduction on group manifolds}
       \end{center}
    \end{bf}   
    \end{Large}
\vspace{0.7cm}

\begin{center}
%\begin{bf}
Hikaru Kawai$^{1),2)}$\footnote
            {
e-mail address : 
hkawai(at)gauge.scphys.kyoto-u.ac.jp}, 
Shinji Shimasaki$^{1)}$\footnote
            {
e-mail address : 
shinji(at)gauge.scphys.kyoto-u.ac.jp}
    and
Asato Tsuchiya$^{3)}$\footnote
           {
e-mail address : 
satsuch(at)ipc.shizuoka.ac.jp}\\
%\end{bf}

\vspace{0.7cm}
                    
 $^{1)}$ {\it Department of Physics, Kyoto University, Kyoto 606-8502, Japan}\\
 \vspace{0.3cm}
 $^{2)}$ {\it Theoretical Physics Laboratory, RIKEN, Wako 351-0198, Japan}\\
 \vspace{0.3cm}
 $^{3)}$ {\it Department of Physics, Shizuoka University}\\
 {\it 836 Ohya, Suruga-ku, Shizuoka 422-8529, Japan}
               
\end{center}

\vspace{0.7cm}

\begin{abstract}
\noindent
We show that the large $N$ reduction holds on group manifolds. 
Large $N$ field theories defined on group manifolds are equivalent to some corresponding matrix models.
For instance, gauge theories on $S^3$ can be regularized in a gauge invariant and $SO(4)$ invariant manner.
% To view the reduced models as bi-local field theories, we reinterpret the large $N$ reduction in real space.
% We also discuss the large $N$ reduction in coset spaces.
\end{abstract}
\vfill
\end{titlepage}
\vfil\eject

\setcounter{footnote}{0}

%%%%%%%%%%%%%%%%%%%%%%%%%%%%%%%%%%%%%%%%%%%%%%%%%%%%%%%%%%%%%%%%%%%%%%%%%%%%%%%%%%%%%%%%%%%%%%%%%%%%%%%%%%%
%%%%%%%%%%%%%%%%%%%%%%%%%%%%%%%%%%%%%%%%%%%%%%%%%%%%%%%%%%%%%%%%%%%%%%%%%%%%%%%%%%%%%%%%%%%%%%%%%%%%%%%%%%%
\section{Introduction}
%%%%%%%%%%%%%%%%%%%%%%%%%%%%%%%%%%%%%%%%%%%%%%%%%%%%%%%%%%%%%%%%%%%%%%%%%%%%%%%%%%%%%%%%%%%%%%%%%%%%%%%%%%%
%%%%%%%%%%%%%%%%%%%%%%%%%%%%%%%%%%%%%%%%%%%%%%%%%%%%%%%%%%%%%%%%%%%%%%%%%%%%%%%%%%%%%%%%%%%%%%%%%%%%%%%%%%%
It has been widely recognized that space-time can be emergent from the degrees of freedom of matrices.
Such emergent space-time was first observed in the large $N$ reduction \cite{EK} (for further developments,
see \cite{Bhanot:1982sh,Parisi:1982gp,Gross:1982at,Das:1982ux,GonzalezArroyo:1982hz,Kazakov:1982zr,
Narayanan:2003fc,Kovtun:2007py, Vairinhos:2007qz,Bringoltz:2009kb,Poppitz:2009fm}).
It asserts that the planar ('t Hooft) limit of gauge theories can be described by
the matrix models obtained by the dimensional reduction to lower (zero) dimensions.
These matrix models are called the (large $N$) reduced models.
%For recent developments in the large $N$ reduction, see \cite{}. 
The large $N$ reduction 
has been studied so far on flat space-time, 
except for a few cases.
It would be important to investigate whether it also holds on curved space-times.
This is because it would provide insight into the description of curved space-times \cite{Hanada:2005vr}
in the matrix models \cite{BFSS,IKKT} that are conjectured 
to give a nonperturbative formulation of string theory and take the form of the reduced model
of ten-dimensional ${\cal N}=1$ super Yang-Mils theory (SYM). 
Practically, it can also be applied to a nonperturbative regularization of planar
gauge theories on curved space-time. 

In this paper, we show that the large $N$ reduction holds on group manifolds, which are typical examples of
curved spaces. In the literature, the mechanism of the large $N$ reduction is  
usually explained in the momentum space. Here we first review it in the real space. 
We see that the reduced model can be viewed as a bi-local field theory with a special feature.
This point of view makes it easy to generalize the large $N$ reduction on flat space to that on group manifolds.
We study the large $N$ reduction for scalar theories in detail. It turns out that
the generalization to gauge theories is straightforward. As an example, we describe the large $N$ reduction for
${\cal N}=4$ SYM on $R\times S^3$. We discuss a relation of
a recently proposed large $N$ reduction for ${\cal N}=4$ SYM on $R\times S^3$ 
\cite{Ishii:2008ib}\footnote{For further developments, see 
\cite{Ishiki:2008te,Ishiki:2009sg,Kitazawa:2008mx,Ishiki:2009vr,Hanada:2009hd,Hanada:2009kz}.} with our version.
We also discuss the large $N$ reduction on coset spaces.

This paper is organized as follows. In section 2, we review the large $N$ reduction for scalar theories on
flat space. We show that the large $N$ reduction holds for the scalar theories on group manifolds in
section 3, and for gauge theories on group manifolds in section 4. 
In section 5, the results in sections 3 and 4 are applied to ${\cal N}=4$ SYM on $R\times S^3$.
Section 6 is devoted to summary and discussion.

%%%%%%%%%%%%%%%%%%%%%%%%%%%%%%%%%%%%%%%%%%%%%%%%%%%%%%%%%%%%%%%%%%%%%%%%%%%%%%%%%%%%%%%%%%%%%%%%%%%%%%%%%%%
%%%%%%%%%%%%%%%%%%%%%%%%%%%%%%%%%%%%%%%%%%%%%%%%%%%%%%%%%%%%%%%%%%%%%%%%%%%%%%%%%%%%%%%%%%%%%%%%%%%%%%%%%%%
\section{Large $N$ reduction on flat space}
%%%%%%%%%%%%%%%%%%%%%%%%%%%%%%%%%%%%%%%%%%%%%%%%%%%%%%%%%%%%%%%%%%%%%%%%%%%%%%%%%%%%%%%%%%%%%%%%%%%%%%%%%%%
%%%%%%%%%%%%%%%%%%%%%%%%%%%%%%%%%%%%%%%%%%%%%%%%%%%%%%%%%%%%%%%%%%%%%%%%%%%%%%%%%%%%%%%%%%%%%%%%%%%%%%%%%%%
To illustrate the large $N$ reduction \cite{EK} on flat space, we consider the scalar $\phi^3$ theory on $R^d$.
The action is given by
\begin{align}
S=\int d^dx \ \Tr \left(\frac{1}{2}(\del_\mu \phi(x))^2+\frac{1}{2}m^2\phi(x)^2
+\frac{1}{3}\kappa \phi(x)^3\right),
\label{phi^3 theory}
\end{align}
where $\phi(x)$ is an $N\times N$ hermitian matrix.
We take the planar ('t Hooft) limit in which 
\begin{align}
N\rightarrow\infty,\;\;\kappa\rightarrow 0 \;\;\;
\mbox{with} \;\; \kappa^2N=\lambda \;\; \mbox{fixed},
\label{'t Hooft limit}
\end{align}
where $\lambda$ is the 't Hooft coupling.

The propagator takes the form
\begin{align}
\langle \phi(x_1)_{ij}\phi(x_2)_{kl}\rangle=D(x_1-x_2)\delta_{il}\delta_{jk}.
%\;\;\; \mbox{with} \;\;
%D(x)=\int\frac{d^dp}{(2\pi)^d}\frac{e^{ipx}}{p^2+m^2}.
\label{D}
\end{align}
The detailed form of $D(x)$ is irrelevant in our argument.
As an example, we calculate the free energy at the two-loop level. There are two 1PI diagrams depicted
in Fig. 1 and Fig. 2.
The diagram in Fig. 1 is planar while the one in Fig. 2 is non-planar.
The result of the planar diagram in Fig. 1 is
\begin{align}
\mbox{Fig. 1}=\frac{1}{6}N^2\lambda\int d^dx_1d^dx_2 \ D(x_1-x_2)^3.
\label{planar contribution in field theory}
\end{align}
The result of the non-planar diagram in Fig. 2 equals that in Fig. 1 divided by $N^2$.
This is an illustration of the well-known fact that only the planar contribution survives in the large $N$ limit.

\begin{figure}[tbp]
\begin{center}
% \psfrag{aa}{$x^1 \!\! ,\, x^2 \!\! ,\, x^3$}
% \psfrag{ab}{$x^0$}
% \psfrag{ba}{$x^1 \!\! ,\, x^2 \!\! ,\, x^3$}
% \psfrag{bb}{$x^0$}
% \psfrag{ca}{$X^M(\sigma^i)$}
\includegraphics[height=10cm, keepaspectratio, clip]{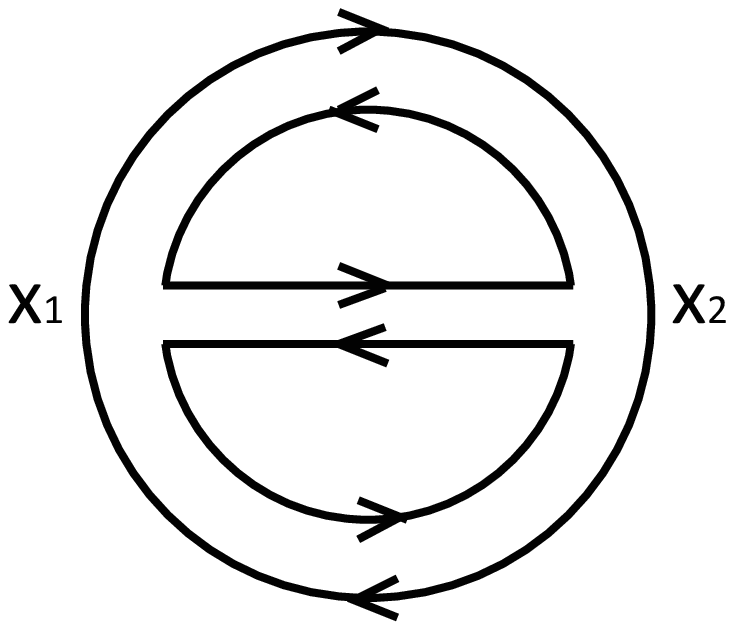}
\end{center}
\vspace{-6cm}
\caption{A planar diagram for the free energy of the scalar $\phi^3$ theory}
\label{planar diagram}
%\end{figure}
%\begin{figure}[tbp]
\begin{center}
\vspace{-1cm}
% \psfrag{aa}{$x^1 \!\! ,\, x^2 \!\! ,\, x^3$}
% \psfrag{ab}{$x^0$}
% \psfrag{ba}{$x^1 \!\! ,\, x^2 \!\! ,\, x^3$}
% \psfrag{bb}{$x^0$}
% \psfrag{ca}{$X^M(\sigma^i)$}
\includegraphics[height=10cm, keepaspectratio, clip]{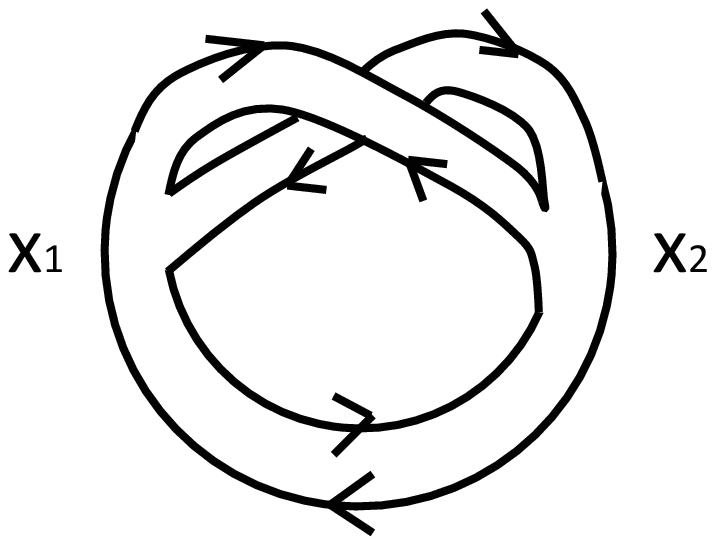}
\end{center}
\vspace{-6cm}
\caption{A non-planar diagram for the free energy of the scalar $\phi^3$ theory}
\label{non-planar diagram}
\end{figure}

% \begin{figure}[tbp]
% \begin{center}
% \begin{minipage}[b]{0.49\textwidth}
%    \begin{center}
%   \includegraphics[width=12cm]{planar.eps}\\
%   \vspace{-5cm}
%    \caption{A planar diagram for the free energy.}
%    \label{planar diagram}
%    \end{center}
% \end{minipage}
%\hspace*{1cm}   
% \begin{minipage}[b]{0.49\textwidth}
%    \begin{center}
%    \includegraphics[width=12cm]{nonplanar.eps}\\
%    \vspace{-5cm}
%    \caption{A non-planar diagram for the free energy.}
%    \label{non-planar diagram}
%    \end{center}
% \end{minipage}   
% \end{center}
%\caption{solution}
%\label{fig:label-2}
% \end{figure}

In order to define the reduced model of (\ref{phi^3 theory}), we consider the space of functions on $R^d$.
The rule to obtain the reduced model is given by
\begin{align}
\phi(x)\rightarrow \hat{\phi},\;\;\; \partial_{\mu}\rightarrow [i\hat{P}_{\mu},\;], \;\;\; \int d^dx \rightarrow v,%\;\;\;\tr\rightarrow \Tr,
\label{rule}
\end{align}
where $\hat{\phi}$ is a hermitian operator acting on the space of function on $R^d$, and 
$\hat{P}_\mu$ is the momentum operator 
which acts on the coordinate basis $|x\rangle \ (x\in R^d)$ as
\begin{align}
 \hat{P}_\mu | x \rangle = -\frac{1}{i}\frac{\partial}{\partial x^\mu}|x \rangle,\;\;\;\;\;
 \langle x|\hat{P}_{\mu}=\frac{1}{i}\frac{\partial}{\partial x^{\mu}}\langle x|.
\label{P_mu}
\end{align}
$v$ is a parameter to be determined later.
%and $\Tr$ is the trace taken over the space of function on $R^d$.
Then, by applying (\ref{rule}) to (\ref{phi^3 theory}), we obtain the reduced model\footnote{While $v$ can be 
absorbed into renormalization of $\kappa$ and $\hat{\phi}$, it turns out that the present normalization is convenient for our argument.}
\begin{align}
S_r=v\Tr\left( \frac{1}{2}[i\hat{P}_\mu,\hat{\phi}]^2+\frac{1}{2}m^2\hat{\phi}^2
+\frac{1}{3}\kappa \hat{\phi}^3 \right),
\label{reduced model of phi^3 theory}
\end{align}
where $\Tr$ is the trace taken over the space of functions on $R^d$.
%At first sight,
(\ref{reduced model of phi^3 theory}) may look different from the reduced model. 
However, it reduces to the familiar form 
if one introduces a momentum cutoff $\Lambda$ and truncates the space of functions on $R^d$
to an $N$-dimensional vector space.
Here we set
\begin{align}
v=\left(\frac{2\pi}{\Lambda}\right)^d,
\end{align}
and take a basis which diagonalizes $\hat{P}_{\mu}$.
Then, $\hat{\phi}$ becomes an $N\times N$ hermitian matrix, and
$\hat{P}_{\mu}$ become constant diagonal matrices whose eigenvalues distribute uniformly in a 
box defined by $-\Lambda/2\leq p_{\mu} \leq \Lambda/2$ in the $d$-dimensional momentum space.
$\Tr$ is viewed as the trace over $N\times N$ matrices.
The introduction of $\Lambda$ and $N$ is interpreted in the real space as follows.
The real space is coarse grained to $N$ $d$-dimensional cubic cells with size $2\pi/\Lambda$. 
This indicates that the volume of the real space
is given by $V=Nv$.

%In what follows,
We reinterpret the large $N$ reduction in the real space,
which makes it easy to generalize the large $N$ reduction on flat space to that on group manifolds.
We denote the matrix element of $\hat{\phi}$ in the coordinate basis
by $\langle x|\hat{\phi}|x'\rangle \equiv \phi(x,x')$, which is
a bi-local field on $R^d$. 
The hermiticity of $\hat{\phi}$ requires that $\phi^*(x,x')=\phi(x',x)$.
Using (\ref{P_mu}), we express (\ref{reduced model of phi^3 theory}) in the coordinate basis as
\begin{align}
 S_r&=v\int d^dx d^dx' \left( 
 -\frac{1}{2}
%\left(\frac{\del}{\del x^\mu}+\frac{\del}{\del {x'}^\mu}\right)
\phi(x',x)\left(\frac{\del}{\del x^\mu}+\frac{\del}{\del {x'}^\mu}\right)^2\phi(x,x')
 +\frac{1}{2}m^2\phi(x',x)\phi(x,x')
 \right) \n
  &\qquad +v\int d^dx d^dx' d^dx'' \ \frac{1}{3} \kappa_r \phi(x,x')\phi(x',x'')\phi(x'',x).
\end{align}
Thus the reduced model can be viewed as a bi-local field theory.
We make a change of variables given by
\begin{align}
X^{\mu}=x^{\mu},\;\;\xi^{\mu}=x^{\mu}-x'{}^{\mu},
\label{change of variables}
\end{align}
and regard $\phi(x,x')$ as a function of $X$ and $\xi$.
$X^{\mu}$ are coordinates of one of the two end-points and $\xi^{\mu}$ are relative coordinates of
the two end-points.
Then, we obtain an equality
\begin{align}
\left(\frac{\del}{\del x^\mu}+\frac{\del}{\del {x'}^\mu}\right)\phi(x,x')
=\frac{\partial}{\partial X^{\mu}}\phi(x,x').
\label{equality}
\end{align}
We see from the equality that the propagator in the reduced model takes the form
\begin{align}
\langle\phi(x_1,x_1')\phi(x_2',x_2)\rangle=\frac{1}{v}D(x_1-x_2)\delta^d((x_1-x_1')-(x_2-x_2')).
\label{propagator in reduced model}
\end{align}
Each end-point propagates as a particle in the original field theory (\ref{phi^3 theory}), while
the relative coordinates are conserved during the propagation.
This implies that
\begin{align}
x_1-x_2=x_1'-x_2'
\label{parallel transport}
\end{align}
in the propagation,
which also follows from the delta function in (\ref{propagator in reduced model}).
In other words, the two end-points are parallely transported.

\begin{figure}[tbp]
\begin{center}
\vspace{-3cm}
% \psfrag{aa}{$x^1 \!\! ,\, x^2 \!\! ,\, x^3$}
% \psfrag{ab}{$x^0$}
% \psfrag{ba}{$x^1 \!\! ,\, x^2 \!\! ,\, x^3$}
% \psfrag{bb}{$x^0$}
% \psfrag{ca}{$X^M(\sigma^i)$}
\includegraphics[height=10cm, keepaspectratio, clip]{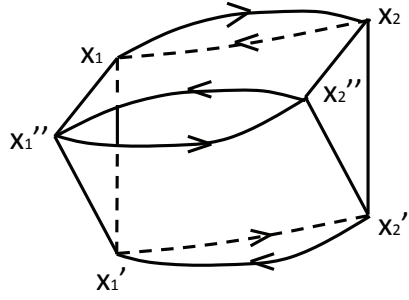}
\end{center}
\vspace{-4cm}
\caption{A planar diagram  for the free energy of the reduced model.}
\label{planar diagram}
\end{figure}

\begin{figure}[t]
\vspace{-1cm}
\begin{center}
% \psfrag{aa}{$x^1 \!\! ,\, x^2 \!\! ,\, x^3$}
% \psfrag{ab}{$x^0$}
% \psfrag{ba}{$x^1 \!\! ,\, x^2 \!\! ,\, x^3$}
% \psfrag{bb}{$x^0$}
% \psfrag{ca}{$X^M(\sigma^i)$}
\includegraphics[height=10cm, keepaspectratio, clip]{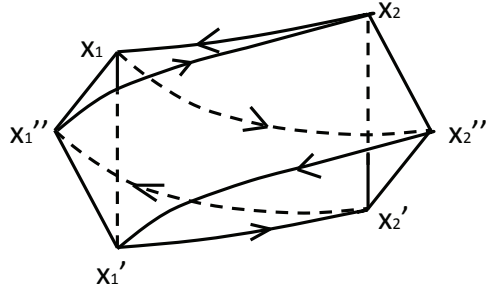}
\end{center}
\vspace{-4cm}
\caption{A non-planar diagram  for the free energy of the reduced model.}
\label{planar diagram}
\end{figure}

Each diagram in the reduced model has the counterpart in the field theory, and vice versa.
As an example, we calculate the free energy of the reduced model at the two-loop level again.
The diagrams in Fig. 3 and Fig. 4 are the counterparts of the diagrams in Fig. 1 and Fig. 2, respectively. 
In Figs. 3 and 4, the aforementioned property of the propagator is visualized.
Here the diagrams in the reduced model that are the counterparts of the planar diagrams in the field theory are still
called the planar diagrams, although they can no longer be drawn on plane.
Similarly, the diagrams in the reduced model that are the counterparts of the non-planar diagrams in
the field theory are called the non-planar diagrams. 

The calculation of the diagram in Fig. 3 is as follows:
\begin{align}
\mbox{Fig. 3}=&\frac{\kappa^2}{6v}\int d^dx_1d^dx_1'd^dx_1''d^dx_2d^dx_2'd^dx_2'' \  D(x_1-x_2)\delta^d((x_1-x_1')-(x_2-x_2')) \n
&\qquad \times D(x'_1-x_2')\delta^d((x_1'-x_1'')-(x_2'-x_2''))D(x_1''-x_2'')\delta^d((x_1''-x_1)-(x_2''-x_2)) \n
=& \frac{\kappa^2}{6v}\delta^d(0)V^2\int d^dx_1d^dx_2 \ D(x_1-x_2)^3.
\label{planar contribution in reduced model}
\end{align}
Indeed, the result can be understood from Fig. 3. We first fix $x_1$ and $x_2$.
Because the relative coordinates are conserved, we have $x_1-x_1'=x_2-x_2'$, and thus 
fixing $x_1'$ implies fixing $x_2'$. Similarly, because of the equation $x_1'-x_1''=x_2'-x_2''$,
fixing $x_1''$ implies fixing $x_2''$. Then, the equation $x_1''-x_1=x_2''-x_2$ yields the factor $\delta^d(0)$.
Fig. 3 shows that $x_1-x_2=x_1'-x_2'=x_1''-x_2''$, which also follows from (\ref{parallel transport}). 
Thus we obtain $\int d^dx_1d^dx_2 D(x_1-x_2)^3$. 
The factor $V^2$ arises from the freedom of $x_1'$ and $x_1''$. The factor $1/v$ comes from the propagators and the vertices.

By comparing (\ref{planar contribution in field theory}) and (\ref{planar contribution in reduced model}) and
using $\delta^d(0)=1/v$ and $V=Nv$, we find that 
the result of the diagram in Fig. 1 divided by $N^2V$ equals that in Fig. 3 divided by $N^2v$
in the limit in which $N\rightarrow\infty$, $v\rightarrow 0$ and $V=Nv\rightarrow\infty$.
It is easy to see that this correspondence holds for all the planar diagrams.

The calculation of the diagram Fig. 4 is as follows: 
\begin{align}
\mbox{Fig. 4}=&\frac{\kappa^2}{6v}\int d^dx_1d^dx_1'd^dx_1''d^dx_2d^dx_2'd^dx_2'' \  D(x_1-x_2)\delta^d((x_1-x_1')-(x_2-x_2')) \n
&\qquad \times D(x_1'-x_2'')\delta^d((x_1'-x_1'')-(x_2''-x_2))D(x_1''-x_2')\delta^d((x_1''-x_1)-(x_2'-x_2'')) \n
&=\frac{\kappa^2}{6v}\delta^d(0)\int d^dx_1d^dx_1'd^dx_2d^dx_2'' D(x_1-x_2)D(x_1'-x_2'')D(x_1-x_2'').
%&=\frac{1}{6}\kappa_r^2\delta^d(0)V\left(\frac{1}{m^2}\right)^3,
\label{non-planar contribution in reduced model}
\end{align}
In this case, $x_1-x_2$, $x_1'-x_2''$ and $x_1''-x_2'$ are all different.
Thus there is no correspondence between the diagrams in Fig. 2 and Fig. 4.
%The above correspondence does not hold between the non-planar diagrams in Fig.2 and Fig. 3.
However, we see from 
(\ref{planar contribution in reduced model}) and (\ref{non-planar contribution in reduced model})
that the result of the diagram in Fig. 4 is suppressed by $1/V^2$ compared with that in Fig. 3 
in the $V\rightarrow\infty$ limit. 

It is easy to verify that in the reduced model all of 
the non-planar diagrams are
suppressed compared with the planar diagrams in the $V\rightarrow\infty$ limit.
Note also that all the non-planar contributions 
are suppressed in the field theory in the large $N$ limit.
We, therefore, find that a relation between the free energy of the field theory $F$ and that of the reduced model $F_r$,
\begin{align}
\frac{F}{N^2V}=\frac{F_r}{N^2v},
\label{relation between free energies}
\end{align}
holds in the limit in which 
\begin{align}
N\rightarrow\infty, \;\; \kappa\rightarrow 0,\;\;v\rightarrow 0 \;\;\;\mbox{with}\;\;
V=Nv \rightarrow\infty, \;\; \lambda=\kappa^2N \;\;\mbox{fixed}.
\label{limit for reduced model in R^d}
\end{align}
It is also easy to see that a relation between the correlation functions,
\begin{align}
\frac{1}{N^{q/2+1}}\langle\Tr(\phi(x_1)\phi(x_2)\cdots\phi(x_q))\rangle
=\frac{1}{N^{q/2+1}}\langle\Tr(\hat{\phi}(x_1)\hat{\phi}(x_2)\cdots\hat{\phi}(x_q))\rangle_r,
\label{relation between correlation functions}
\end{align}
holds in the limit (\ref{limit for reduced model in R^d}),
where $\langle\cdots\rangle$ and $\langle\cdots\rangle_r$ denote the expectation values in the field theory and the reduced model, respectively,
and $\hat{\phi}(x)$ is defined by
\begin{align}
\hat{\phi}(x)=e^{i\hat{P}_{\mu}x^{\mu}}\hat{\phi}e^{-i\hat{P}_{\nu}x^{\nu}}.
\label{hatphix}
\end{align}
Thus the reduced model retrieves the planar limit of the original field theory.

We close this section with a comment on the large $N$ reduction on $T^d$ with a finite volume $V$.
In this case, the above suppression for the non-planar diagrams in the reduced model 
no longer exists. To resolve this problem, we modify the reduced model
as follows. We introduce a ultraviolet momentum cutoff $2\pi/v^{1/d}$ such that 
\begin{align}
v=V/n
\end{align}
with an integer $n$.
This can also be interpreted as dividing the real space into $n$ $d$-dimensional
cubic cells such that the volume of each cell
is given by $v$. The space of functions on $T^d$ is expressed as an $n$-dimensional vector space.
We consider a tensor product space of this vector space and a $k$-dimensional vector space and put $N=nk$, which is
nothing but the dimension of the tensor product space.
We make the operator $\hat{\phi}$ act on the tensor product space. Equivalently, we make $\phi(x,x')$ carry
extra matrix indices:
\begin{align}
\phi(x,x') \rightarrow \phi(x,x')_{\alpha\beta} \;\;\;(\alpha,\beta=1,\cdots, k).
\end{align}
In (\ref{rule}) and (\ref{reduced model of phi^3 theory}), we replace $\hat{P}_{\mu}$ by
$\hat{P}_{\mu}\otimes 1_k$ 
and regard $\Tr$ as the trace taken over the tensor product space.
%In what follows, we often omit $\otimes 1_k$ for economy of notation.
All of the equations below (\ref{reduced model of phi^3 theory}) are changed according to the above recipe.
In the reduced model, we take a limit in which
\begin{align}
n\rightarrow\infty, \;\; k\rightarrow\infty, \;\; \kappa\rightarrow 0,\;\;\;\mbox{with}\;\;
\lambda=\kappa^2N=\kappa^2nk \;\;\mbox{fixed}.
\label{limit for reduced model in T^d}
\end{align}
Then, the non-planar diagrams are suppressed at least by $1/k^2$ compared with the planar diagrams.
It is easy to verify that (\ref{relation between free energies}) and (\ref{relation between correlation functions}) with
$\hat{\phi}(x)=e^{i\hat{P}_{\mu}x^{\mu}\otimes 1_k}\hat{\phi}e^{-i\hat{P}_{\nu}x^{\nu}\otimes 1_k}$ still hold in the limit
(\ref{limit for reduced model in T^d}),
so that the reduced model retrieves the planar limit of
the original field theory.
Note that $T^d$ can be identified with $U(1)^d$, which is a compact connected Lie group. 
In the next section, the result for
$U(1)^d$ in this section is generalized to general compact connected Lie groups.

%%%%%%%%%%%%%%%%%%%%%%%%%%%%%%%%%%%%%%%%%%%%%%%%%%%%%%%%%%%%%%%%%%%%%%%%%%%%%%%%%%%%%%%%%%%%%%%%%%%%%%%%%%%
%%%%%%%%%%%%%%%%%%%%%%%%%%%%%%%%%%%%%%%%%%%%%%%%%%%%%%%%%%%%%%%%%%%%%%%%%%%%%%%%%%%%%%%%%%%%%%%%%%%%%%%%%%%
\section{Large $N$ reduction on group manifolds}
%%%%%%%%%%%%%%%%%%%%%%%%%%%%%%%%%%%%%%%%%%%%%%%%%%%%%%%%%%%%%%%%%%%%%%%%%%%%%%%%%%%%%%%%%%%%%%%%%%%%%%%%%%%
%%%%%%%%%%%%%%%%%%%%%%%%%%%%%%%%%%%%%%%%%%%%%%%%%%%%%%%%%%%%%%%%%%%%%%%%%%%%%%%%%%%%%%%%%%%%%%%%%%%%%%%%%%%
In this section, we study the large $N$ reduction on group manifolds. 
It turns out that the argument runs parallel to the case of flat space in the previous section.

%First, we describe some properties of group manifolds.
Let $G$ be a compact connected Lie group and $t_a \;\;(a=1,\cdots, \mbox{dim}\:G)$ be generators of its Lie algebra.
$t_a$ satisfy a commutation relation $[t_a,t_b]=if_{ab}^{\;\;\;c}t_c$.
We consider a space of functions on $G$, where the coordinate basis are denoted by $|g\rangle\;\;(g\in G)$.
For $h\in G$, the left translation in $G$ is expressed as
\begin{align}
\hat{U}_L(h)|g\rangle=|hg\rangle, \;\;\;
\langle g|\hat{U}_L(h)=\langle h^{-1}g|,
\end{align}
while the right translation in $G$
\begin{align}
\hat{U}_R(h)|g\rangle=|gh^{-1}\rangle, \;\;\;
\langle g|\hat{U}_R(h)=\langle gh|.
\end{align}
A function on $G$, $\psi(g)=\langle g|\psi\rangle$, is transformed under the above
translations as
\begin{align}
&(\hat{U}_L(h)\psi)(g)=\langle g| \hat{U}_L(h) |\psi\rangle=\langle h^{-1}g|\psi\rangle=\psi(h^{-1}g), \n
&(\hat{U}_R(h)\psi)(g)=\langle g| \hat{U}_R(h)|\psi\rangle=\langle gh|\psi\rangle=\psi(gh).
\label{left and right translations}
\end{align}
We define the generators of the left (right) translation, $\hat{L}_a$ ($\hat{K}_a$), in terms of infinitesimal translations generated by $e^{i\epsilon t_a}$ as
\begin{align}
e^{i\epsilon \hat{L}_a}=\hat{U}_L(e^{i\epsilon t_a}), \;\;\;
e^{i\epsilon \hat{K}_a}=\hat{U}_R(e^{i\epsilon t_a}).
\end{align}
Using the commutation relation for $t_a$, it is easy to see that
\begin{align}
[\hat{L}_a,\hat{L}_b]=if_{ab}^{\;\;\;c}\hat{L}_c,\;\;\;
[\hat{K}_a,\hat{K}_b]=if_{ab}^{\;\;\;c}\hat{K}_c,\;\;\;
[\hat{L}_a,\hat{K}_b]=0.
\label{commutators}
\end{align}
% or equivalently
% \begin{align}
% &\hat{U}_R(h^{-1})\hat{L}_a\hat{U}_R(h)=0, \;\;\; \hat{U}_L(h^{-1})\hat{L}_a\hat{U}_L(h)=R_{ab}(h^{-1})\hat{L}_b, \n
% &\hat{U}_L(h^{-1})\hat{K}_a\hat{U}_L(h)=0, \;\;\; \hat{U}_R(h^{-1})\hat{K}_a\hat{U}_R(h)=R_{ab}(h^{-1})\hat{K}_b, 
% \label{ajoint action}
% \end{align}
% where $R_{ab}(h)$ is the representation matrix of the adjoint representation of $G$, which is an orthogonal matrix.
$\hat{L}_a$ ($\hat{K}_a$) is the right (left) invariant Killing vector.
$\hat{L}_a$ and $\hat{K}_a$ act on functions on $G$ as differential operators,
which we denote by ${\cal L}_a$ and ${\cal K}_a$,
respectively:
\begin{align}
&\hat{L}_a|g\rangle=-{\cal L}_a|g\rangle,\;\;\;\; \langle g|\hat{L}_a={\cal L}_a\langle g|, \n
&\hat{K}_a|g\rangle=-{\cal K}_a|g\rangle,\;\;\;\; \langle g|\hat{K}_a={\cal K}_a\langle g|,
\label{L_a and K_a}
\end{align}
which are analogous to (\ref{P_mu}).
%${\cal L}_a$ and ${\cal K}_a$ obey the same algebra as (\ref{commutators}).
We define the right invariant 1-forms $e^a$ and the left invariant 1-forms $s^a$ by
\begin{align}
d=dx^{\mu}\frac{\partial}{\partial x^{\mu}}=ie^a{\cal L}_a=is^a{\cal K}_a,
\label{invariant 1-forms}
\end{align}
where $x^{\mu}\;\;(\mu=1\cdots, \mbox{dim}\;G)$ are coordinates parameterizing $G$.  
It follows from (\ref{commutators}) that the invariant 1-forms satisfy 
the Maurer-Cartan equations
\begin{align}
de^a-\frac{1}{2}f_{bc}^{\;\;\;a}e^b\wedge e^c=0,\;\;\; ds^a-\frac{1}{2}f_{bc}^{\;\;\;a}s^b\wedge s^c=0.
\label{Maurer-Cartan equation}
\end{align}
The left and right invariant metric $h_{\mu\nu}$ is defined in terms of $e^a$ or $s^a$ by\footnote{In general, the invariant metric can be
defined by any invariant rank-2 symmetric tensor. Because we can assume that $\delta_{ab}$ is such a tensor, 
we use it for simplicity.}
\begin{align}
h_{\mu\nu}=e^a_{\mu}e^a_{\nu}=s^a_{\mu}s^a_{\nu}.
\end{align}
The Haar measure of $G$ is given by
\begin{align}
dg=e^1\wedge e^2\wedge \cdots  \wedge e^{\mbox{\scriptsize{dim}}\:G},
\label{Haar measure}
\end{align}
and the volume of the manifold is given by $V=\int dg$, which is finite.
% \begin{align}
% dgg^{-1}=-ie^at_a,\;\;\; g^{-1}dg=is^at_a.
% \end{align}
% which are dual to the right and left invariant Killing vectors, respectively, in the sense that

We consider the scalar $\phi^3$ theory on $G$. Noting that $h^{\mu\nu}\partial_{\mu}\phi\partial_{\nu}\phi=-({\cal L}_a\phi)^2$, 
we can write down the action as\footnote{Here
higher derivative kinetic terms can also be considered.}
\begin{align}
S=\int dg \ \Tr\left( -\frac{1}{2}(\cL_a\phi(g))^2+\frac{1}{2}m^2\phi(g)^2
+\frac{1}{3}\kappa \phi(g)^3 \right),
\label{phi^3 theory on group manifold}
\end{align}
$\phi(g)$ is an $N\times N$ hermitian matrix whose elements are
functions on $G$.
%$dg$ is the Haar measure of $G$, and the volume of the manifold is given by
%$V=\int dg$. 
The theory possesses the $G\times G$ symmetry. Namely, 
it is invariant under the transformations, $\phi'(g)=(\hat{U}_L(h)\phi)(g)$ 
and $\phi'(g)=(\hat{U}_R(h)\phi)(g)$.
We take the planar ('t Hooft) limit (\ref{'t Hooft limit}).
The propagator takes the form
\begin{align}
\langle\phi(g_1)_{ij}\phi(g_2)_{kl}\rangle=\Delta(g_1g_2^{-1})\delta_{il}\delta_{jk}.
\end{align}
The detailed form of $\Delta(g)$ is again irrelevant in our argument.
% The diagrams in (\ref{phi^3 theory on group manifold}) agree with those in
% (\ref{phi^3 theory}). For instance, the value of the diagram in Fig.1 (with $x$ and $y$ replaced by
% $g_1$ and $g_2$) is
% \begin{align}
% \mbox{Fig.1}=\frac{1}{6}N^2\lambda\int dg_1dg_2 \ \Delta(g_1g_2^{-1})^3,
% \label{planar contribution in field theory on group manifold}
% \end{align}

We define the reduced model of (\ref{phi^3 theory on group manifold}) as follows.
As in the case of $T^d$, we consider the tensor product space of the space of functions on $G$ and a $k$-dimensional vector space.
%Let $\hat{\phi}$ be a hermitian operator acting on the tensor product space.
The rule to obtain the reduced model on $G$, which is analogous to (\ref{rule}),  is 
\begin{align}
\phi(g)\rightarrow \hat{\phi},\;\;\; {\cal L}_a\rightarrow [\hat{L}_a\otimes 1_k,\;], \;\;\; \int dg \rightarrow v,%\;\;\;\tr\rightarrow \Tr,
\label{rule on group manifold}
\end{align}
where $\hat{\phi}$ is a hermitian operator acting on the tensor product space, 
and $\Tr$ is the trace taken over the tensor product space.
In what follows, we often omit $\otimes 1_k$ for economy of notation.
Applying (\ref{rule on group manifold}) to (\ref{phi^3 theory on group manifold}), we obtain the reduced model 
\begin{align}
S_r=v\Tr\left( -\frac{1}{2}[\hat{L}_a,\hat{\phi}]^2+\frac{1}{2}m^2\hat{\phi}^2
+\frac{1}{3}\kappa\hat{\phi}^3 \right).
\label{matrix phi3 G}
\end{align}
%where $\Tr$ is taken over the space of functions on $G$. 
%We see from (\ref{ajoint action}) that 
The reduced model also possesses the $G\times G$ symmetry given by
\begin{align}
\hat{\phi}'= \hat{U}_L(h_L)\hat{U}_R(h_R)\hat{\phi}\hat{U}_R(h_R^{-1})\hat{U}_L(h_L^{-1}).
\end{align}
We express the action in terms of the coordinate basis.
We denote the matrix element of $\hat{\phi}$ by $\langle g|\hat{\phi}|g'\rangle \equiv \phi(g,g')$, which is
a bi-local $k\times k$ matrix field on $G$.
The hermiticity of $\hat{\phi}$ is translated into the relation $\phi(g,g')^{\dagger}=\phi(g',g)$.
Then, using (\ref{L_a and K_a}), (\ref{matrix phi3 G}) is expressed as
\begin{align}
S_r&=v\int dg dg' \tr\left\{
 \frac{1}{2}\phi(g',g)\left(\cL_a^{(g)}+\cL_a^{(g')}\right)^2\phi(g,g')
 +\frac{1}{2}m^2\phi(g',g)\phi(g,g')
 \right\} \n
 &\qquad +v\int dg dg' dg'' \ \frac{1}{3}\kappa \ \tr(\phi(g,g')\phi(g',g'')\phi(g'',g)),
\end{align}
where $\tr$ is the trace over $k\times k$ matrices.
The reduced model is again viewed as a bi-local field theory on $G$.
We make a change of variables, which is a counterpart of (\ref{change of variables}),
\begin{align}
u=g,\;\; \zeta=g'{}^{-1}g,
\label{change of variables in reduced model}
\end{align}
and regard $\phi(g,g')$ as a function of $u$ and $\zeta$. Noting that $\zeta$ is invariant under
the left translation, we find an equality
\begin{align}
\left(\cL_a^{(g)}+\cL_a^{(g')}\right)\phi(g,g')=\cL_a^{(u)}\phi(g,g').
\label{equality in reduced model}
\end{align}
Note also that the Haar measures are invariant under the change of variables (\ref{change of variables in reduced model}).
It follows from this fact and the equality (\ref{equality in reduced model}) that the propagator in the reduced model takes the form
\begin{align}
\langle\phi(g_1,g_1')_{\alpha\beta}\phi(g_2',g_2)_{\gamma\delta}\rangle
=\frac{1}{v}\Delta(g_1g_2^{-1})\delta(g_1'{}^{-1}g_1,g_2'^{-1}g_2)\delta_{\alpha\delta}\delta_{\beta\gamma},
\label{propagator in reduced model on group manifold}
\end{align}
where $\alpha,\;\beta,\;\gamma,\;\delta=1,\cdots,k$, and $\delta(g_1,g_2)$ is the delta function under the Haar measure, which satisfies 
\begin{align}
\delta(g_1,g_2)=\delta(g_2,g_1)=\delta(hg_1,hg_2)=\delta(g_1h,g_2h)
\label{property of delta function}
\end{align}
for arbitrary $h\in G$. (\ref{propagator in reduced model on group manifold})
is a counterpart of (\ref{propagator in reduced model}) and indicates
that the propagator in the reduced model on $G$ has the same property as the one on flat space.

Because of the form of the propagator (\ref{propagator in reduced model on group manifold}) and
the property of the delta function (\ref{property of delta function}), the calculation of the diagrams
in the reduced model on $G$ proceeds in the same manner as that on flat space.
Therefore, we find that the large $N$ reduction holds on $G$.

We now consider the ultraviolet regularization.
The space of functions on $G$ is identified with the representation space $V_{reg}$ 
of the regular representation of $G$.
The elements of $G$ act on the representation space as (\ref{left and right translations}).
%For a compact Lie group $G$, 
$V_{reg}$ has the following decomposition as a vector space\footnote{This follows from the Peter-Weyl theorem. It states that
a function on $G$, $\psi(g)$, can be expanded as $\psi(g)=\sum_r\sum_{ij}c_{ij}^{[r]}R^{[r]}_{ij}(g)$, where $R^{[r]}_{ij}(g)$ is the representation matrix for
the irreducible representation $r$.},
\begin{align}
V_{reg}=\bigoplus_rV_{r^*}\otimes V_r,
\label{irreducible decomposition of regular representation}
\end{align}
where $r$ labels the irreducible representations, $r^*$ denotes the complex conjugate
representation of $r$, and $V_r$ is the representation space of the representation $r$.
%and $d_r$ is the dimension of the representation $r$. 
The left translation acts on the left $V_{r^*}$, while the right translation on the right $V_r$.
Namely, $\hat{L}_a$ and $\hat{K}_a$ act on (\ref{irreducible decomposition of regular representation}) as
\begin{align}
&\hat{L}_a=\bigoplus_r L^{[r]}_a \otimes 1_{d_r}, \n
&\hat{K}_a=\bigoplus_r 1_{d_r}\otimes L^{[r]}_a,
\label{form of hatLa}
\end{align}
where $L^{[r]}_a$ are the representation matrices of $t_a$ in the representation $r$, 
and $d_r$ is the dimension of the representation $r$.

To regularize the theory, we first consider the set of irreducible representations $I_{\Lambda}$ for 
a positive number $\Lambda$ given by
\begin{align}
I_{\Lambda}=\{r;C_2(r)<\Lambda^2\},
\end{align}
where $C_2(r)$ is the second-order Casimir of the representation $r$.
We then restrict the range of the sums in (\ref{irreducible decomposition of regular representation}) and
(\ref{form of hatLa}) to $I_{\Lambda}$,
and put $n=\sum_{r\in I_{\Lambda}}d_r^2$ and $v=V/n$. The $\Lambda\rightarrow\infty$ limit
corresponds to the $n\rightarrow\infty$ limit, and
$\Lambda$ plays the role of the ultraviolet cutoff.
Thus the space of functions on $G$ is truncated to an $n$-dimensional vector space.
$\hat{L}_a$ in (\ref{matrix phi3 G}) is explicitly given by
\begin{align}
\left(\bigoplus_{r\in I_{\Lambda}} L^{[r]}_a \otimes 1_{d_r}\right)\otimes 1_k.
\label{matrix background}
\end{align}
% The reduced model is expressed as
% \begin{align}
% S_r=\Tr \left(-\frac{1}{2}[L_a,\phi]^2+\frac{1}{2}m^2\phi^2+\frac{1}{3}\kappa\phi^3\right),
% \end{align}
% where $\phi$ is an $N\times N$ matrix.
It is remarkable that the $G\times G$ symmetry
is preserved even after the above ultraviolet regularization is introduced.
We take the limit given in (\ref{limit for reduced model in T^d}). Then, the relation (\ref{relation between free energies}) holds.
The counterpart of (\ref{relation between correlation functions}),
\begin{align}
\frac{1}{N^{q/2+1}}\langle\Tr(\phi(g_1)\phi(g_2)\cdots\phi(g_q))\rangle
=\frac{1}{N^{q/2+1}}\langle\Tr(\hat{\phi}(g_1)\hat{\phi}(g_2)\cdots\hat{\phi}(g_q))\rangle_r,
\label{relation between correlation functions on group manifold}
\end{align}
also holds in the limit (\ref{limit for reduced model in T^d}), where 
$\hat{\phi}(g)$ is defined by
\begin{align}
\hat{\phi}(g)=e^{i\theta^a\hat{L}_a}\hat{\phi}e^{-i\theta^b\hat{L}_b},
\label{hatphix}
\end{align}
for $g=e^{i\theta^at_a}$.
Thus the reduced model (\ref{matrix phi3 G}) retrieves the planar limit of 
the original field theory on $G$ (\ref{phi^3 theory on group manifold}).

%%%%%%%%%%%%%%%%%%%%%%%%%%%%%%%%%%%%%%%%%%%%%%%%%%%%%%%%%%%%%%%%%%%%%%%%%%%%%%%%%%%%%%%%%%%%%%%%%%%%%%%%%%%
%%%%%%%%%%%%%%%%%%%%%%%%%%%%%%%%%%%%%%%%%%%%%%%%%%%%%%%%%%%%%%%%%%%%%%%%%%%%%%%%%%%%%%%%%%%%%%%%%%%%%%%%%%%
\section{Gauge theory on group manifold}
%%%%%%%%%%%%%%%%%%%%%%%%%%%%%%%%%%%%%%%%%%%%%%%%%%%%%%%%%%%%%%%%%%%%%%%%%%%%%%%%%%%%%%%%%%%%%%%%%%%%%%%%%%%
%%%%%%%%%%%%%%%%%%%%%%%%%%%%%%%%%%%%%%%%%%%%%%%%%%%%%%%%%%%%%%%%%%%%%%%%%%%%%%%%%%%%%%%%%%%%%%%%%%%%%%%%%%%
In this section, we extend the large $N$ reduction on group manifolds found in the previous section to the case of gauge theories.

First, we consider the reduced model of Yang-Mills (YM) theory  on a group manifold $G$, which is compact and connected.
% We define the right invariant 1-form $e^a$ by
% \begin{align}
% idgg^{-1}=e^at_a.
% \end{align}
% It is dual to the right-invariant Killing vector in the sense that
% \begin{align}
% d=ie^a{\cal L}_a.
% \label{e^a and calL_a}
% \end{align}
% It is easy to show that it satisfies the Maurer-Cartan equation
% \begin{align}
% de^a-\frac{1}{2}f_{bc}^{\;\;\;a}e^b\wedge e^c=0.
% \label{Maurer-Cartan equation}
% \end{align}
We write down $U(N)$ YM theory on $G$ in a form directly connected to the reduced model.
We expand the gauge field $A$, which is an $N\times N$ hermitian matrix, in terms of $e^a$ as 
\begin{align}
A=X_ae^a.
\end{align}
Then, the field strength is expressed as
\begin{align}
F&=dA+iA\wedge A \n
%&=dX_ae^a+X_ade^a+iX_aX_be^a\wedge e^b \n
&=\frac{1}{2}(i{\cal L}_aX_b-i{\cal L}_bX_a+f_{ab}^{\;\;\;c}X_c+i[X_a,X_b])e^a\wedge e^b,
\label{F}
\end{align}
where (\ref{invariant 1-forms}) and (\ref{Maurer-Cartan equation}) has been used to obtain the second line.
The term $f_{ab}^{\;\;\;c}X_c$ in (\ref{F}) comes from the curvature and in general makes the gauge field massive.
Using (\ref{F}) and (\ref{Haar measure}), $U(N)$ YM theory is rewritten 
in terms of ${\cal L}_a$ and $X_a$ as\cite{Ishii:2008tm}
\begin{align}
S&=\frac{1}{4\kappa^2}\int \Tr(F\wedge \ast F) \n
&=-\frac{1}{4\kappa^2}\int dg \;\Tr({\cal L}_aX_b-{\cal L}_bX_a-if_{ab}^{\;\;\;c}X_c+[X_a,X_b])^2.
\label{YM on G}
\end{align}

Applying the rule (\ref{rule on group manifold}) to (\ref{YM on G}), we obtain the reduced model of YM theory on $G$
\begin{align}
S_r&=-\frac{v}{4\kappa^2}\Tr ([\hat{L}_a,\hat{X}_b]-[\hat{L}_b,\hat{X}_a]-if_{ab}^{\;\;\;c}\hat{X}_c+[\hat{X}_a,\hat{X}_b])^2 \n
&=-\frac{v}{4\kappa^2}\Tr ([\hat{L}_a+\hat{X}_a,\hat{L}_b+\hat{X}_b]-if_{ab}^{\;\;\;c}(\hat{L}_c+\hat{X}_c))^2,
\label{reduced model of YM}
\end{align}
where we have used (\ref{commutators}) to obtain the second line.
Repeating the argument in the previous section, we find that if the limit (\ref{limit for reduced model in T^d})
is taken, the reduced model (\ref{reduced model of YM}) retrieves the planar limit of 
the original YM theory (\ref{YM on G}), 
aside from a possible problem discussed below. 
The relation (\ref{relation between free energies}) holds, and it is easy to obtain from 
(\ref{relation between correlation functions on group manifold}) a relation between the expectation values of the Wilson loops\footnote{The same type of the Wilson loop in RHS of (\ref{relation between Wilson loops})
is studied in \cite{Ishiki:2009vr}.}
\begin{align}
&\left\langle \frac{1}{N}\Tr\left(P\exp\left[i\int_c A_{\mu}(x(\sigma))\frac{dx^{\mu}(\sigma)}{d\sigma}d\sigma \right]\right)\right\rangle \n
&=\left\langle \frac{1}{N}\Tr\left(P\exp\left[i\int_c(\hat{L}_a+\hat{X}_a) 
e^a_{\mu}(x(\sigma))\frac{dx^{\mu}(\sigma)}{d\sigma}d\sigma\right]\right)\right\rangle_r,
\label{relation between Wilson loops}
\end{align}
where $C$ stands for a closed path parametrized by $\sigma$ on $G$.

Remarkably, the second line in (\ref{reduced model of YM}) indicates that
redefining $\hat{X}_a$ as $\hat{L}_a+\hat{X}_a\rightarrow \hat{X}_a$, 
namely absorbing $\hat{L}_a$ into $\hat{X}_a$, leads to
\begin{align}
S_r'=-\frac{v}{4\kappa^2}\Tr\left([\hat{X}_a,\hat{X}_b]-if_{ab}^{\;\;\;c}\hat{X}_c\right)^2,
\label{reduced model of YM 2}
\end{align}
which is nothing but the dimensionally reduced model of (\ref{YM on G}) to zero dimension.
Similarly, the redefinition makes the Wilson loop in RHS of (\ref{relation between Wilson loops}) the dimensional reduction of that
in LHS.
This is the original idea of the large $N$ reduction. That is, the planar limit of YM theory
is described by a matrix that is obtained by the dimensional
reduction to zero dimension.
Indeed, the redefinition is rephrased as follows.
(\ref{reduced model of YM}) is the theory obtained 
by expanding (\ref{reduced model of YM 2}) around 
a classical solution $\hat{X}_a=\hat{L}_a$ of (\ref{reduced model of YM 2}). 
The gauge symmetry of the original YM theory corresponds to
the symmetry of the reduced model given by
\begin{align}
\hat{X}'_a=\hat{U}\hat{X}_a\hat{U}^{\dagger},
\end{align}
where $\hat{U}$ is an arbitrary unitary operator.
Thus the reduced model can give a regularization that preserves the gauge symmetry. 

In the case of YM theory on flat space, the same absorption also happens, where
the classical backgrounds are given by $\hat{P}_{\mu}$. However, these backgrounds are unstable against the quantum correction 
due to the massless modes. 
This instability is interpreted as the so-called $U(1)^d$ symmetry breaking \cite{Bhanot:1982sh}. 
We need remedy such as the quenching \cite{Bhanot:1982sh,Gross:1982at} 
for the reduced model to reproduce the original theory. 
In our case, if $G$ is semi-simple, the theory
(\ref{reduced model of YM}) is massive, so that the background $\hat{L}_a$ is stable to all order in
the coupling constant. Furthermore, the tunneling to other classical solutions is suppressed in the large $k$ limit. 
Hence, we can just expand (\ref{reduced model of YM 2}) around $\hat{L}_a$ without any remedy.
This is advantageous in the large $N$ reduction of supersymmetric gauge theories, because the quenching is not compatible with supersymmetry. In our case, the reduced model preserves supersymmetries that
the background $\hat{L}_a$ preserves among those of the original field theory.
%In the next section, as an example, we will discuss the large $N$ reduction for ${\cal N}=4$ SYM theory on $R\times %S^3$, where $S^3$ is identified with the $SU(2)$ group manifold. 
If $G$ is not semi-simple, we need remedy such as the quenching.

Next, we consider the large $N$ reduction of a fermion in the adjoint representation. 
The action of the fermion on $G$ is
\begin{align}
S&=-\frac{1}{\kappa^2}
\int dg \: \Tr\left(\bar{\psi}\gamma^ae^{\mu}_a\left(\partial_{\mu}\psi+\frac{1}{4}\omega_{\mu}^{bc}\gamma_{bc}\psi+i[A_{\mu},\psi]\right)+m\bar{\psi}\psi\right),
\label{fermion action on group manifold}
\end{align}
where the spin connection is determined by the equation
\begin{align}
de^a+\omega^a_{\;\;b}\wedge e^b=0.
\label{torsionless condition}
\end{align}
Comparing (\ref{torsionless condition}) with (\ref{Maurer-Cartan equation}), we find
\begin{align}
\omega^a_{\;\;b}=\frac{1}{2}f_{bc}^{\;\;\;a}e^c.
\label{spin connection}
\end{align}
Substituting (\ref{spin connection}) into (\ref{fermion action on group manifold})
and using $e_a^{\mu}A_{\mu}=X_a$, we obtain
\begin{align}
S=-\frac{1}{\kappa^2}\int dg \: \Tr\left(\bar{\psi}\gamma^a(i{\cal L}_a\psi+i[X_a,\psi])+\frac{1}{8}f_{abc}\bar{\psi}\gamma^{abc}\psi+m\bar{\psi}\psi\right).
\label{fermion action on group manifold 2}
\end{align}
Note that the third term in (\ref{fermion action on group manifold 2}) is a mass term coming from the curvature.
Applying the rule (\ref{rule on group manifold}) to (\ref{fermion action on group manifold 2}) yields the reduced model of the fermion on $G$
\begin{align}
S_r=-\frac{v}{\kappa^2}\Tr \left(i\bar{\hat{\psi}}\gamma^a[\hat{L}_a+\hat{X}_a,\hat{\psi}]
+\frac{1}{8}f_{abc}\bar{\hat{\psi}}\gamma^{abc}\hat{\psi}+m\bar{\hat{\psi}}\hat{\psi}\right).
\end{align}
The redefinition $\hat{L}_a+\hat{X}_a\rightarrow \hat{X}_a$ again leads to the dimensionally reduced model of
(\ref{fermion action on group manifold 2}). 
It is remarkable that there is no fermion doublers in the reduced model
unlike the fermion on the lattice.

The same absorption of the background $\hat{L}_a$ occurs in the
case of scalar fields in the adjoint representation.
We, therefore, conclude that if $G$ is semi-simple, the planar limit of a gauge theory on $G$ with the matter fields in the adjoint representation
is equivalent to the theory obtained by expanding its dimensionally reduced model
around a classical solution $\hat{L}_a$.
The reduced model preserves the gauge symmetry, the $G\times G$ symmetry (and (part of) supersymmetries)
of the original (supersymmetric) gauge theory.
If $G$ is not semi-simple, remedy such as the quenching is needed for the large $N$ reduction to hold.

% Next, we discuss the large $N$ reduction of a scalar field in the adjoint representation. The coupling of the scalar field to the gauge field arise from a
% replacement in the kinetic term
% \begin{align}
% i{\cal L}_a\phi \rightarrow i{\cal L}_a\phi+i[X_a,\phi].
% \end{align}
% Applying the rule (\ref{}) to the RHS gives

%%%%%%%%%%%%%%%%%%%%%%%%%%%%%%%%%%%%%%%%%%%%%%%%%%%%%%%%%%%%%%%%%%%%%%%%%%%%%%%%%%%%%%%%%%%%%%%%%%%%%%%%%%%
%%%%%%%%%%%%%%%%%%%%%%%%%%%%%%%%%%%%%%%%%%%%%%%%%%%%%%%%%%%%%%%%%%%%%%%%%%%%%%%%%%%%%%%%%%%%%%%%%%%%%%%%%%%
\section{${\cal N}=4$ SYM on $R\times S^3$: an example}
%%%%%%%%%%%%%%%%%%%%%%%%%%%%%%%%%%%%%%%%%%%%%%%%%%%%%%%%%%%%%%%%%%%%%%%%%%%%%%%%%%%%%%%%%%%%%%%%%%%%%%%%%%%
%%%%%%%%%%%%%%%%%%%%%%%%%%%%%%%%%%%%%%%%%%%%%%%%%%%%%%%%%%%%%%%%%%%%%%%%%%%%%%%%%%%%%%%%%%%%%%%%%%%%%%%%%%%
In this section, we apply the results in sections 3 and 4 to ${\cal N}=4$ SYM on $R\times S^3$.
This theory has a superconformal symmetry $SU(2,2|4)$, whose algebra includes thirty-two supercharges, 
and is equivalent to ${\cal N}=4$ SYM on $R^4$ through a conformal mapping.
Its reduced model can serve as a nonperturbative formulation of planar ${\cal N}=4$ SYM, which would be important
in the study of the AdS/CFT correspondence. 

We regard $S^3$ as the $SU(2)$ group manifold.  
The isometry of $S^3$, $SO(4)=SU(2)\times SU(2)$, corresponds to the left and right translations.
The elements of $SU(2)$ are parametrized in terms of the Euler angles as
\begin{equation}
g=e^{-i\varphi \sigma_3/2}e^{-i\theta \sigma_2/2}e^{-i\psi \sigma_3/2},
\label{Euler angles}
\end{equation}
where $\sigma_a \;\;(a=1,2,3)$ are the Pauli matrices, 
and $0\leq \theta\leq \pi$, $0\leq \varphi < 2\pi$, $0\leq \psi < 4\pi$.
The right invariant 1-forms are given by
\begin{eqnarray}
&&e^1=-\sin \varphi d\theta + \sin\theta\cos\varphi d\psi,\nonumber\\
&&e^2=\cos \varphi d\theta + \sin\theta\sin\varphi d\psi,\nonumber\\
&&e^3=d\varphi + \cos\theta d\psi,
\end{eqnarray}
which satisfy the Maurer-Cartan equation $de^a-\frac{1}{2}\epsilon_{abc}e^b\wedge e^c=0$. The right invariant Killing vector is given by
\begin{eqnarray}
&&{\cal{L}}_1=-i\left(-\sin\varphi\partial_{\theta}-\cot\theta\cos\varphi\partial_{\varphi}+\frac{\cos\varphi}{\sin\theta}\partial_{\psi}\right),\nonumber\\
&&{\cal{L}}_2=-i\left(\cos\varphi\partial_{\theta}-\cot\theta\sin\varphi\partial_{\varphi}+\frac{\sin\varphi}{\sin\theta}\partial_{\psi}\right),\nonumber\\
&&{\cal{L}}_3=-i\partial_{\varphi},
\label{Killing vector}
\end{eqnarray}
which satisfy the commutation relation $[{\cal L}_a,{\cal L}_b]=i\epsilon_{abc}{\cal L}_c$.
The invariant metric is given by
\begin{equation}
ds^2=e^ae^a=
d\theta^2+\sin^2\theta d\varphi^2 +(d\psi+\cos\theta d\varphi)^2.
\label{metric of S^3}
\end{equation}
We have fixed the radius of $S^3$ to 2. The Haar measure is given by $dg=\sin\theta d\theta d\varphi d\psi$, and
$V=16\pi^2$.

% The action of ${\cal N}=4$ SYM on $R\times S^3$ is 
% \begin{align}
% S=\frac{1}{\kappa^2}\int dg \left(-\frac{1}{4}F_{\mu\nu}F^{\mu\nu}-\frac{1}{2}D_{\mu}X_mD^{\mu}X_m-\frac{1}{2}X_m^2
% -\frac{i}{2}\bar{\lambda}\Gamma^0D_0\lambda-\frac{1}{2}\bar{\lambda}\Gamma^a e_{a}^{\mu}\right)
% \end{align}

The action of ${\cal N}=4$ SYM on $R\times S^3$ is given in ten-dimensional notation by
\begin{align}
S=\frac{1}{4\kappa^2}\int dtdg \ \Tr
&\left(\frac{1}{4}F_{\hat{\mu}\hat{\nu}}F^{\hat{\mu}\hat{\nu}}+\frac{1}{2}D_{\hat{\mu}}X_mD^{\hat{\mu}}X_m
+\frac{1}{8}X_m^2-\frac{1}{4}[X_m,X_n]^2 \right.\n
&\left.+\frac{1}{2}\Psi^{\dagger}D_t\Psi+\frac{i}{2}\Psi^{\dagger}\gamma^ae_a^{\mu}D_{\mu}\Psi
-\frac{1}{2}\Psi^{\dagger}\gamma^m[X_m,\Psi]\right),
\end{align}
where $\mu=\theta,\varphi,\psi$ while $\hat{\mu},\hat{\nu}=t,\theta,\varphi,\psi$, and $m,n=4,\cdots,9$.
The covariant derivatives are defined by 
$D_{\hat{\mu}}=\nabla_{\hat{\mu}}+i[A_{\hat{\mu}},\;]$,
where $\nabla_{\mu}$ include the spin connection for the fermion.
The mass term for the adjoint scalars $X_m$ comes from the coupling of the conformal scalars to
the scalar curvature of $S^3$.
We apply the dimensional reduction we studied for general group manifolds in the previous section to
${\cal N}=4$ SYM on $R\times S^3$ to obtain a theory on $R$.
The resulting action takes the form of 
the plane wave matrix model (PWMM) \cite{Berenstein:2002jq}, which
was first pointed out in \cite{Kim:2003rza} (see also \cite{Lin:2005nh}).
Thus the reduced model of ${\cal N}=4$ SYM on $R\times S^3$ is given by
\begin{align}
S_r
%_{\rm pp}
= \frac{v}{\kappa^2}
\int
dt \, \mbox{Tr} 
&\left[ \frac{1}{2}(D_tX_M)^2-\frac{1}{4}[X_M,X_N]^2 
+\frac{1}{2}\Psi^{\dagger} D_t \Psi
-\frac{1}{2}\Psi^{\dagger}\gamma^M[X_M,\Psi] \right.\nonumber\\
&\;\;\left.+\frac{1}{2}(X_a)^2 
+\frac{1}{8}(X_m)^2 +i\epsilon_{abc}X_aX_bX_c
+\frac{3i}{8}\Psi^{\dagger}\gamma^{123}\Psi \right],
\label{PWMM}
\end{align}
where $M,N$ run from 1 to 9.
%$1 \le M,N \le 9$, $1 \le a,b,c \le 3$ and $4 \le m \le 9$.
%The covariant derivative is defined by
%$D_t=\partial_t+i[A, \;]$.
$A_t$, $X_M$ and $\Psi$ are $N\times N$ matrices depending on $t$. We have omitted the hats on these matrices.
%The mass term for $X_m$ comes from the coupling of the conformal scalar to the scalar curvature of $S^3$. 
The mass term for $X_a$ and the Myers term arises from the $F^2$ term, while the mass term for $\Psi$ from the spin connection. The model (\ref{PWMM}) possesses the $SU(2|4)$ symmetry, which is a subgroup of $SU(2,2|4)$ and whose algebra
includes sixteen supercharges.

Any classical solutions of (\ref{PWMM}) in which $X_a$ are given by
a reducible representation of $SU(2)$ preserve the $SU(2|4)$ symmetry. 
Following (\ref{matrix background}), we pick up the following solution and
expand (\ref{PWMM}) around it:
\begin{align}
L_a
=\left(\begin{array}{cccc}
L_a^{[0]} &&& \\
& L_a^{[1/2]}\otimes 1_{2} && \\
&&\ddots & \\
&&& L_a^{[K]}\otimes 1_{2K+1} 
\end{array} \right) \otimes 1_k,
\label{S^3 background}
\end{align}
where $L_a^{[j]}$ are the representation matrices of the $SU(2)$ generators in the spin $j$ representation.
The background (\ref{S^3 background}) recovers the $SO(4)$ symmetry, which is the isometry
of $S^3$, as mentioned around (\ref{matrix background}).
%Namely, the theory expanded around this background has an enhanced symmetry $SU(2)\times SU(2|4)$.
The reduced model (\ref{PWMM}) gives a regularization that respects 
at least the $SU(2)\times SU(2|4)$ symmetry and the gauge symmetry.
%$SO(4)$ and $SU(2|4)$ are subgroups of $SU(2,2|4)$, and
%the algebra of the latter includes sixteen supercharges. 
It follows that $\Lambda\simeq K$, $n=\sum_{j=0}^K(2j+1)^2$ and $N=nk$. Then, the reduced model (\ref{PWMM}) retrieves 
planar ${\cal N}=4$ SYM on $R\times S^3$ in the limit (\ref{limit for reduced model in T^d}).

% On the other hand, it is shown in \cite{Ishii:2008ib} that the theory around 
% a classical solution of (\ref{PWMM}),
On the other hand, in \cite{Ishii:2008ib}, the following classical solution of (\ref{PWMM}) is considered:
\begin{align}
L_a
=\left(\begin{array}{cccc}
L_a^{[N_0/2-T/4-1/2]}&&& \\
& L_a^{[N_0/2-T/4]} && \\
&&\ddots & \\
&&& L_a^{[N_0/2+T/4-1/2]} 
\end{array} \right) \otimes 1_k,
\label{S^3 background 2}
\end{align}
where $N_0$ and $T$ are a positive integer and a positive even integer, respectively. $N$ and $v$ are expressed in terms of $N_0,\;T,\;k$ as
\begin{align}
&N=(T+1)N_0k, \n
&v=\frac{16\pi^2}{(T+1)N_0^2}.
\end{align}
It is shown in \cite{Ishii:2008ib} that the theory around (\ref{S^3 background 2})
is equivalent to planar ${\cal N}=4$ SYM on $R\times S^3$ in the limit in which
\begin{align}
&\kappa\rightarrow 0,\;\; N_0\rightarrow\infty,\;\;T\rightarrow\infty, \;\;k\rightarrow\infty \n
&\mbox{with}\;\; \lambda=\kappa^2N\;\;\mbox{fixed and}\;\;T/N_0\rightarrow 0.
\label{S^3 limit}
\end{align}
The statement can be viewed as another large $N$ reduction for ${\cal N}=4$ SYM on $R\times S^3$, and
has passed some nontrivial tests \cite{Ishiki:2008te,Ishiki:2009sg,Kitazawa:2008mx}.
This background preserves the $SU(2|4)$ symmetry, so that this type of the large $N$ reduction
gives a regularization that preserves the $SU(2|4)$ symmetry and the gauge symmetry.
Here $S^3$ is viewed as an $S^1$-bundle over $S^2$.
$T$ corresponds to the ultraviolet cutoff for the Kaluza-Klein momentum along $S^1$, while
$N_0$ to that for the Kaluza-Klein momentum on $S^2$. 
The two models defined around 
the two backgrounds (\ref{S^3 background}) and (\ref{S^3 background 2}) of PWMM
belong to the same universality class.
Indeed, we can show that the perturbative expansion around
(\ref{S^3 background}) in the limit (\ref{limit for reduced model in T^d})
eventually agrees with that around (\ref{S^3 background 2}) in the limit (\ref{S^3 limit}).
Remarkably, the two models can be put on a computer 
in terms of the method \cite{Anagnostopoulos:2007fw,Catterall:2008yz} to study the strongly coupled regime
of ${\cal N}=4$ SYM.

%%%%%%%%%%%%%%%%%%%%%%%%%%%%%%%%%%%%%%%%%%%%%%%%%%%%%%%%%%%%%%%%%%%%%%%%%%%%%%%%%%%%%%%%%%%%%%%%%%%%%%%%%%%
%%%%%%%%%%%%%%%%%%%%%%%%%%%%%%%%%%%%%%%%%%%%%%%%%%%%%%%%%%%%%%%%%%%%%%%%%%%%%%%%%%%%%%%%%%%%%%%%%%%%%%%%%%%
\section{Summary and discussion}
%%%%%%%%%%%%%%%%%%%%%%%%%%%%%%%%%%%%%%%%%%%%%%%%%%%%%%%%%%%%%%%%%%%%%%%%%%%%%%%%%%%%%%%%%%%%%%%%%%%%%%%%%%%
%%%%%%%%%%%%%%%%%%%%%%%%%%%%%%%%%%%%%%%%%%%%%%%%%%%%%%%%%%%%%%%%%%%%%%%%%%%%%%%%%%%%%%%%%%%%%%%%%%%%%%%%%%%
In this paper, we showed that the large $N$ reduction holds on group manifolds.
As an example, we described the large $N$ reduction for ${\cal N}=4$ SYM on $R\times S^3$.
%We revealed a relation of this example with the recent proposal made in \cite{}.

While we studied YM theories in sections 4 and 5, we can consider a Chern-Simons-like theory on $G$
defined by
\begin{align}
S=\frac{1}{\omega^2}\int dg f^{abc}\Tr\left(iX_a{\cal L}_bX_c+\frac{1}{2}f_{bc}^{\;\;\;d}X_aX_d
+\frac{2i}{3}X_aX_bX_c\right),
\label{CS theory}
\end{align}
which has the $G\times G$ symmetry and the $U(N)$ gauge symmetry.
Here $\omega$ is determined by the invariance of $e^{iS}$ under the large gauge transformations.
For $G=SU(2)$, this agrees with pure Chern-Simons theory on $S^3$.
The reduced model of (\ref{CS theory}) is given by
\begin{align}
S_r=\frac{v}{\omega^2}f^{abc}\Tr\left(\frac{1}{2}f_{bc}^{\;\;\;d}\hat{X}_a\hat{X}_d
+\frac{2i}{3}\hat{X}_a\hat{X}_b\hat{X}_c\right).
\label{reduced model of CS theory}
\end{align}
Repeating the arguments in sections 3 and 4, we can show that expanded around (\ref{matrix background}),
the reduced model (\ref{reduced model of CS theory}) 
retrieves the original theory (\ref{CS theory}) in the limit 
(\ref{limit for reduced model in T^d}). 
%For instance, we can apply this result to pure Chern-Simons theory and ABJM theory on $S^3$. 
In this manner, the large $N$ reduction holds for
a wide class of gauge theories including ones in the Veneziano limit,
quiver gauge theories \cite{Kazakov:1982zr} and the ABJM theory \cite{Aharony:2008ug}.
The details of the study of (\ref{CS theory}) and (\ref{reduced model of CS theory})
will be reported in \cite{KST}.

The large $N$ reduction on coset spaces $G/H$ is also an interesting problem.
There is a simple prescription to obtain reduced models of scalar theories on coset spaces $G/H$
from the corresponding reduced models on $G$. Let $\hat{R}_A\;\;(A=1,\cdots, \mbox{dim}\;H)$ 
be the generators of the Lie algebra of $H$.
The prescription is to impose a condition $[\hat{R}_A,\hat{\phi}]=0$ for all $A$, or equivalently $\phi(r^{-1}g,r^{-1}g')=\phi(g,g')$ for arbitrary $r\in H$.
This is, for instance, achieved by adding a mass term $M^2\Tr[\hat{R}_A,\hat{\phi}]^2$ with large $M$ to the 
reduced models on $G$.
We will further investigate the case of other theories on $G/H$ \cite{KST}.

We have restricted ourselves to compact connected Lie groups so far.
Indeed, the argument in section 3 still holds formally for the case of non-compact connected Lie groups, where
the $k$-dimensional vector space is not needed because $V$ is infinite.
However, to establish the large $N$ reduction on such group manifolds,
we need to resolve a problem in infrared regularization.
Infinite $V$ gives rise to a continuous spectrum, which is not compatible with finite-size matrices.
As seen in section 2, theories on $R^d$ are obtained by an infinite volume limit of the corresponding theories on $T^d$.
Similarly, a possible resolution of the above problem is to define
a theory on a non-compact group by an infinite volume limit of the corresponding theory on a compact group or 
a coset space.

We hope that our findings in this paper will lead to a progress in 
the problem of describing curved space-times in matrix models conjectured to give
a nonperturbative formulation of superstring.

%%%%%%%%%%%%%%%%%%%%%%%%%%%%%%%%%%%%%%%%%%%%%%%%%%%%%%%%%%%%%%%%%%%%%%%%%%%%%%%%%%%%%%
\section*{Acknowledgment}
%%%%%%%%%%%%%%%%%%%%%%%%%%%%%%%%%%%%%%%%%%%%%%%%%%%%%%%%%%%%%%%%%%%%%%%%%%%%%%%%%%%%%%
This work was supported by the Grant-in-Aid for the Global COE program ``The Next Generation
of Physics, Spun from Universality and Emergence" from the Ministry of Education, Culture,
Sports, Science and Technology (MEXT) of Japan.
The work of S.\ S.\ 
is supported 
%in part 
by JSPS.
%is supported in part by the JSPS Research Fellowship for Young Scientists. 
The work of A.\ T.\ 
is supported 
%in part 
by Grant-in-Aid for Scientific Research (19540294) from JSPS.

\end{document}